\documentclass[lettersize,journal]{IEEEtran}
\IEEEoverridecommandlockouts
\usepackage{cite}
\usepackage{amsmath,amssymb,amsfonts}
\usepackage{mathtools}

\usepackage{cuted}
\usepackage{graphicx}
\usepackage{textcomp}
\usepackage{xcolor}
\usepackage{cuted}
\usepackage{amsmath, amssymb, amsthm,algorithm}
\usepackage[latin1]{inputenc}

\usepackage{latexsym}
\usepackage{graphics,epsfig}
\usepackage{amsfonts}
\usepackage{booktabs}
\usepackage{color}
\usepackage{multirow}
\usepackage[justification=centering]{caption}
\usepackage{dsfont}
\usepackage{graphicx}
\usepackage{subcaption} 
\usepackage{adjustbox}
\usepackage{kantlipsum}
\usepackage[caption=false,font=footnotesize]{subfig}
\usepackage{cases}
\usepackage{url}
\usepackage{makecell}
\usepackage{soul}

\usepackage{array}
\usepackage{algpseudocode}
\usepackage{mathtools,lipsum,cuted}

\usepackage[justification=centering]{caption}

\usepackage[caption=false,font=footnotesize]{subfig}
\usepackage[T1]{fontenc}
\usepackage{mwe}
\usepackage{subfig}
\usepackage{pdflscape}

\newlength{\tempheight}
\newlength{\tempwidth}

\newcommand{\rowname}[1]
{\rotatebox{90}{\makebox[\tempheight][c]{#1}}}

\newcommand{\columnname}[1]
{\makebox[\tempwidth][c]{#1}}

\makeatletter
\def\BState{\State\hskip-\ALG@thistlm}
\makeatother

\newcounter{example}[section]

\usepackage[english]{babel}
\usepackage{verbatim}

\usepackage[english]{babel}

\theoremstyle{plain}

\theoremstyle{remark}

\newcommand\Tstrut{\rule{0pt}{2.6ex}}         
\newcommand\Bstrut{\rule[-0.9ex]{0pt}{0pt}}   

\def\BibTeX{{\rm B\kern-.05em{\sc i\kern-.025em b}\kern-.08em
    T\kern-.1667em\lower.7ex\hbox{E}\kern-.125emX}}
\begin{document}

\voffset=0.05in
\textheight=9.28in

\title{A Cross-Layer Analysis of Network Antifragility with RIS-assisted Links under Jamming Attacks}

\author{\IEEEauthorblockN{Mounir Bensalem, Thomas R\"othig and Admela Jukan}\\
\IEEEauthorblockA{Technische Universit\"at Braunschweig, Germany;
\{mounir.bensalem, t.roethig,  a.jukan\}@tu-bs.de}

}

\maketitle

\begin{abstract}
Antifragility is an economics term defined as measure of (monetary) benefits gained from the adverse events and variability of the markets. This paper integrates for the first time the antifragility into the network based on communication links with Reconfigurable Intelligent Surface (RIS) affected by a jamming attack. We analyze whether antifragility can be achieved for several jamming models. Beyond the link-level gains, the results reveal how antifragile RIS-assisted links can be integrated into multi-hop systems to improve end-to-end network resilience, connectivity, and throughput under adversarial effects.
\end{abstract}


\section{Introduction}

\IEEEPARstart{C}{onventional} resilient network systems aim to preserve or restore functionality under hostile conditions. By contrast, antifragility, as defined in economics, leverages uncertainty and adverse conditions to achieve performance gains\cite{Taleb.2013}. In networking, antifragility has remained largely unexplored to date. With the emergence of multi-hop wireless links based on Reconfigurable Intelligent Surfaces (RIS), antifragility may become a salient feature in networks,  due to the RIS property of using two-dimensional arrays of tunable reflecting elements to manipulate signal amplitude and phase, mitigating interference and extending coverage \cite{9430683}.

This paper proposes an analytical model of a multi-hop RIS-assisted network system, including a transmitter, a receiver, a RIS and 2 types of jammers,  and derives antifragile performance gains of the network under jamming attacks. We analyze various jamming models on links, including Digital Radio Frequency Memory (DRFM) and phase and amplitude shifting. By leveraging the RIS reflections, we prove that the network can improve throughput under certain conditions, demonstrating a remarkably antifragile behavior. Specifically, when the baseline data rate is low and adversarial power is high, we observe a notable antifragile gain factor of up to $5\times$. Although the gain factor is smaller for larger baseline data rates, it still shows a notable performance gain. This work is the first one to integrate antifragitly of communication links into RIS enabled wireless mobile networks. 

\section{Related Work}

Related work focuses on antifragility in communications, and our study is the first to explore antifragility in wireless networks with RIS in a cross-layer analysis. For link antifragility, paper \cite{Lichtman.2017} presents a method to design frequency-shift keying (FSK) waveforms in order to exploit reactive jammers, effectively forcing an attacker, i.e., the jammer node,for performance gains of legitimate users. Building on this fundamental idea, the early work in \cite{Lichtman.2018} studied the antifragile wireless links. More recently, \cite{Ji.2023}, basing its antifragility scheme on paper \cite{Lichtman.2017} focused on reducing the outage probability in multi-relay networks. We extend the antifragility scope to consider various jamming models, signal separation and spoofing signal creation, which other works do not consider. Regarding the assumptions on link channel model, we adopt the known RIS channel model, e.g., from \cite{9780032,  bjornson2020rayleigh, Bensalem.11620241182024, 10086588}.  The attacker's channel model is adapted from  \cite{Lichtman.2017, 9780032}. Paper \cite{Sun.2021} is related due to a joint beamforming optimization scheme that maximizes both the secrecy rate and the system throughput, reducing the impact of the jamming attack. By contrast, we model the jammer as an information relaying node and configure the RIS phases solely to boost the SNR of legitimate users, i.e., without mechanisms to degrade the eavesdropping link. Paper \cite{tsiota2019jamming} offers a system-level view of jamming attacks, showing other network effects beyond  link-level Tx-Rx-Jammer model. Paper \cite{10066528} integrates beamforming with artificial noise injection to suppress the eavesdroppers. Instead, we assume the jammers signal as a constructive relay without artificial noise injection. 


\section{Physical Layer Assumptions}

\begin{figure}[ht]
\centering
\includegraphics[scale=0.5]{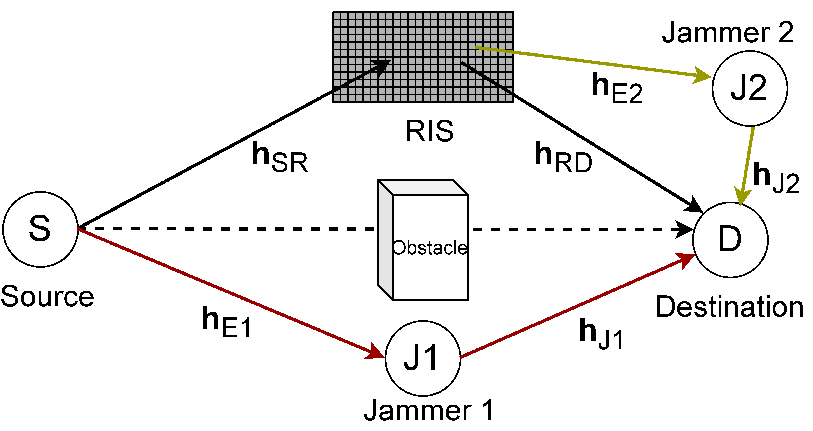}
\caption{Two hop network model with two jamming paths.}
\label{fig:arch}
\end{figure}

The system model is illustrated in Figure \ref{fig:arch}, including source, receiver, one RIS node, and two jammers. Two types of jamming paths are illustrated, with Jammer 1 and Jammer 2 nodes. We refer to the path through Jammer 1 as \emph{Source-Aware Eavesdropping Path}, and to Jammer 2 as \emph{RIS-Aware-Eavesdropping Path}. To model the performance of the two paths in a cross layer analysis, it is necessary to start from the physical layer channel and jammer node models.  

\subsection{Source- and RIS-aware eavesdropping paths model}
For both models, the signal received at the destination node in a jammer free network system is given based on \cite{Bensalem.11620241182024, 9780032,  bjornson2020rayleigh},
\begin{multline}
\label{eq:signal_m1}
y(t) = \mathbf{h}_{SR} \mathbf{R}^{1/2} \Phi \mathbf{R}^{1/2} \mathbf{h}_{RD} x(t) + w(t) \\
    = \left( \sum_{a=1}^{M} \sum_{k=1}^{M} \sum_{l=1}^{M} \rho^{1/2}_{a,k} \rho^{1/2}_{a,l} h_{SR,k} h_{RD,l} e^{j\phi_a} \right) x(t) + w(t)
    \end{multline} 

Where  $M$ denotes the  number of RIS elements, $x(t)$ is the transmitted signal from the source to the destination node, $w(t)$ is the noise,  $\mathbf{h}_{SR} = [h_{SR,1},..,h_{SR,k},...,h_{SR,M}]$ is the fading channel from the source to RIS, with $h_{SR,k} = g_{SR,k}e^{-j\theta_k}\sqrt{(d_{SR})^{-\delta}}$ is the channel coefficient related to the element $k$,   and $\mathbf{h}_{RD}=[h_{RD, 1},...,h_{RD, l},...,h_{RD, M}]^T$ is the fading channel from the RIS to the destination, with   $h_{RD, l} = g_{RD, l}e^{-j\psi_{k,l}}\sqrt{(d_{RD})^{-\delta}}$ is the channel coefficient related to the element $l$,   $\delta$ denotes  the path loss exponent,  $g_{\mathrm{SR, k}}$ and  $g_{\mathrm{RD, l}}$ the channel gains for S-to-R through element $k$ and R-to-D through element $l$,  $d_{\mathrm{SR}}$ and  $d_{\mathrm{RD}}$ the distance from the source to the RIS and RIS to the destination, respectively.  Moreover, the correlation matrix $\mathbf{R}$ is defined as an $M\mathrm{x}M$ matrix, that defines the correlation coefficients $\rho_{i,j}$ between the $i^{th}$ and $j^{th}$ elements of the RIS. Thus, $\forall i,j, 0 \leq \rho_{i,j}\leq 1 $ and $\rho_{i,j} = 1$ for $i = j$. In addition, the reflection coefficients of the RIS are within the diagonal phase matrix $\Phi = \mathrm{diag}[e^{-j\phi_1},e^{-j\phi_2},....,e^{-j\phi_M}]$.

We denote by $\mathbf{h}_{\mathrm{E1}}$ the eavesdropping CSI from the source to Jammer 1, and by $\mathbf{h}_{\mathrm{J1}}$ the jamming CSI from Jammer 1 to destination, which is given by  $h_{\mathrm{s}}$ \cite{Lichtman.2017}: 
\begin{equation}
    h_{\mathrm{s}} = \sqrt{\frac{\kappa}{\kappa + 1}} \sigma e^{j\theta_{s,i}} + \sqrt{\frac{1}{\kappa + 1}} \sum_{i=1}^L R_{s, i} e^{j\theta_{s,i}}, s= J1 \text{ or } E1
\label{eq:eq_jam_channel_model_1}
\end{equation}
Where  $\kappa$ is the Rician factor, $\sigma^2$ the  average power, $L$ the number of paths,  and $R_{s, i}$ the Rayleigh distributed amplitude and $\theta_{s,i}$ the uniformly distributed phase of the channel. When the Rician factor $\kappa=0$, the Rician fading channel coefficients are simplified to the Rayleigh fading coefficients.
 The received signal on the path with Jammer 1 is given by:
\begin{multline}
y(t)
= \mathbf{h}_{\mathrm{SR}}\,\mathbf{R}^{1/2}\,\boldsymbol{\Phi}\,\mathbf{R}^{1/2}\,\mathbf{h}_{\mathrm{RD}}\,x(t) + y_{J1}(t) + w(t)
 \\[1mm]
= \Bigl(\sum_{a=1}^{M}\sum_{k=1}^{M}\sum_{l=1}^{M}
      h_{\mathrm{SR},k}\,h_{\mathrm{RD},l}\,
      \rho_{a,k}^{\tfrac12}\,\rho_{a,l}^{\tfrac12}\,e^{j\phi_a}\Bigr)\,x(t) \\[1mm]
\quad
+ A\Bigl(\sum_{i=1}^{L}R_{\mathrm{E1},i}\,e^{j\theta_{\mathrm{E1},i}}\Bigr)
  \Bigl(\sum_{j=1}^{L}R_{\mathrm{J1},j}\,e^{j\theta_{\mathrm{J1},j}}\Bigr)
  x\bigl(t-\tau_{\mathrm{J1}}\bigr)
  + w(t)
\end{multline}

With $A$ being a multiplicative factor that can be defined based on the type of jamming attack (Table~\ref{tab:jammer_models}, Section \ref{subsubsec:jam}), and $\tau_{\mathrm{J1}}$ denotes the end-to-end path delay.

We denote the eavesdropping CSI from RIS to Jammer 2 by $\mathbf{h}_{\mathrm{RJ}}$ and by $\mathbf{h}_{\mathrm{J2}}$ the jamming CSI from Jammer 2 to the  destination, calculated using Eq. (\ref{eq:eq_jam_channel_model_1}).
The received signal $y(t)$ is a superposition of the legitimate signal defined by Eq. (\ref{eq:signal_m1}), and the jamming signal (Jammer 2), which is defined by $y_{J2}(t)$ and taken from \cite{9780032},
\begin{multline} \label{eq:signal_total}
 y_{\mathrm{J2}}(t)= 
\left( \sum_{a=1}^{M} \sum_{k=1}^{M} \sum_{l=1}^{M} 
\rho_{a,k}^{1/2}\,\rho_{a,l}^{1/2}\, h_{SR,k}\, h_{RJ,l}\, e^{j\phi_a} \right)  \\
 A 
 \left( \sum_{j=1}^{L} R_{\mathrm{J2},j}\,e^{j\theta_{\mathrm{J2},j}} \right) x(t - \tau_{\mathrm{J2}})
\end{multline}
where $\tau_{\mathrm{J2}}$ denotes the delay of the jamming path through Jammer 2.
Thus, the received signal  $y(t)$ is given by:
\begin{multline} \label{eq:signal_total}
y(t) =  \mathbf{h}_{SR} \mathbf{R}^{1/2} \boldsymbol{\Phi} \mathbf{R}^{1/2} \mathbf{h}_{RD} x(t) + y_{\mathrm{J2}}(t) +w(t) \\
= \left( \sum_{a=1}^{M} \sum_{k=1}^{M} \sum_{l=1}^{M} 
\rho_{a,k}^{1/2}\,\rho_{a,l}^{1/2}\, h_{\mathrm{SR},k}\, h_{\mathrm{RD},l}\, e^{j\phi_a} \right)x(t)\\
 + 
\left( \sum_{a=1}^{M} \sum_{k=1}^{M} \sum_{l=1}^{M} 
\rho_{a,k}^{1/2}\,\rho_{a,l}^{1/2}\, h_{SR,k}\, h_{RJ,l}\, e^{j\phi_a} \right)  \\
 A 
 \left( \sum_{j=1}^{L} R_{\mathrm{J2},j}\,e^{j\theta_{\mathrm{J2},j}} \right) x(t - \tau_{\mathrm{J2}}))+ w(t)
\end{multline}

It should be noted that Jammer 2 can also be a malicious RIS, able to make attacks using phase shifting or amplitude shifting, as an active malicious RIS could do. If the destination is equipped with $m$ antennas, the resulting signals become:
\begin{equation}
    \mathbf{Y} = \mathbf{s_{v}}\mathbf{y} + \mathbf{w(t)}
\end{equation}


With $\mathbf{Y} \in \mathbb{C}^{m\mathrm{x}1}$ for a single received symbol, $\mathbf{y} \in \mathbb{C}^{1\mathrm{x}1}$, $\mathbf{s_{v} \in \mathbb{C}^{m\mathrm{x}1}}$ being the steering vector for the angle of arrival (AoA) and $y$ being $y=h_\mathrm{l} x(t) + h_\mathrm{J2} x(t-\tau_{\mathrm{j}})$ for the single antenna and $y=h_\mathrm{l} x(t) + h_\mathrm{J2} x(t)$ for the multipath jammer.

\subsection{Jammer model}\label{subsubsec:jam}
A reactive (repeater) jammer operates by capturing the signal emitted by the legitimate transmitter, optionally applying a deterministic transformation, and retransmitting the modified signal with the objective of degrading the receiver's performance. We analyze three categories of jammer models: digital radio frequency memory (DRFM), phase shifting (PS) and Amplitude shifting (AS), as summarized in Table \ref{tab:jammer_models}. 
 The DRFM attack retransmits the target signal on a sample-by-sample basis with a constant amplification gain denoted as $\beta_a$. Assuming that $h_{\mathrm{E1}}$ and $h_{\mathrm{J1}}$ are defined using Eq. (\ref{eq:eq_jam_channel_model_1}), the DRFM jamming signal can be expressed as follows:
\begin{equation}
    y_J(t) = \beta_a  h_{\mathrm{E1}}h_{\mathrm{J1}}x(t-\tau_{\mathrm{J1}}) + w(t)
\end{equation}

The PS jammer transforms the signal, randomly inverting the phase of the intercepted symbols. We denote by  $U(t)$ a random sequence drawn from the set $U(t) \in \{1, -1\}$ with probability mass function  $f_U (u) = 0.5 : U = 1, -1$. 

\begin{equation}
    y_J(t) = U(t)  h_{\mathrm{E1}}h_{\mathrm{J1}}x(t-\tau_{\mathrm{J1}}) + w(t)
\end{equation}
Similarly, the AS jammer introduces random amplitude perturbations, defined as $V(t)$. Thus the signal is given by:
\begin{equation}
    y_J(t) = V(t)  h_{\mathrm{E1}}h_{\mathrm{J1}}x(t-\tau_{\mathrm{J1}}) + w(t)
\end{equation}

\begin{table}[h!]
\centering
\begin{tabular}{|p{1.9cm}|p{1.1cm}|p{4.5cm}|}
\hline
\Tstrut \Bstrut \textbf{Jammer Model} & \textbf{Factor A}&\textbf{Signal Manipulation} \\ \hline
\Tstrut \Bstrut DRFM & $\beta_a$ & Amplification  using fixed $\beta_a$\\  \hline
\Tstrut \Bstrut PS & $U(t)$ & Phase shift using $U(t): u \in \left\{ 1, -1 \right\}$ \\ \hline
\Tstrut \Bstrut AS & $V(t)$ & Amplitude amplification  attenuation using $V(t): 2 \geq v \geq 0$ \\ \hline
\end{tabular}
\caption{Jammer models summarized \cite{Lichtman.2018}.}
\label{tab:jammer_models}
\end{table}

\section{Antifragility Scheme Design}
In this section, we first estimate the path delay and attack detection. To achieve antifragile gain, the network system must first detect a jamming attack and estimate the path delay with jammers relative to the legitimate path, $\tau_{\mathrm{J1}}$ or $\tau_{\mathrm{J2}}$. Additionally, the signals at the receiver must be received orthogonally, either in time or space, to avoid overlap that would hinder demodulation and decoding. We then provide a jamming classification, and based on signal adaptation, calculate the antifragile gain. The proposed antifragility gain model can be easilyt extended to network routing. For instance, by reporting antifragile gain metrics to the routing system, the system can dynamically update link weights or routing tables to favor links exhibiting performance gains under interference. In dense or distributed RIS deployments, this allows the network to exploit jamming-induced variability as a routing advantage, enabling self-optimizing, interference-aware topologies.

\subsection{Jamming detection and path delay estimation}
The jamming signal can be detected by evaluating the Bit Error Rate (BER) of the received signal. Without loss of generality, we assume that Reed-Solomon (RS) code is used. 
The presence of a jammer is detected when the received BER is such that the RS code is unable to decode the message. When this condition occurs, a Cross-Correlation (CC) analysis is performed on the signal.
The CC metric is obtained by correlating the received sequence $y$ with its time-aligned replica $\tilde{y}$; the presence of a peak is taken as evidence of a jammer. Because the delay estimate $\tau$ provides only coarse alignment-insufficient for symbol- or sample-level precision-we compute a complete cross-correlation (i.e., a sliding inner product) rather than a single dot product, given by \cite{Lichtman.2018}:

\begin{equation} 
R_{y\tilde{y}}(\tau) = \sum_{n=0}^{F_{max}} y[n] \cdot \tilde{y}[n + \tau]
\label{eq:CC}
\end{equation}
where $F_{max}$ is the frame length.
If a malicious signal is present, the CC analysis displays a secondary sharp peak, in dependence on the strength and similarity of the malicious signal to the original signal. This peak is then used as an initial estimate for the delay introduced by the jammer and is defined as the delay that maximizes the magnitude of the cross-correlation:
\begin{equation}
\hat{\tau} = \arg \max_{\tau \in [-\gamma, \gamma ]} |R_{y\tilde{y}}(\tau)|
\end{equation}
where $\gamma$  is maximum anticipated delay value.
Since most jammers operate in cycles, the receiver stores the timing of the first jamming attack and continues the transmission. If a second attack occurs, the system calculates the duration of the jamming cycle and can adapt the legitimate signal.


\subsection{Orthogonality Approach}
\label{subsec:beam}
To ensure orthogonal reception of desired and jamming waveforms, we consider two operating modes: (i) spatial orthogonality via directional separation, and, (ii) temporal orthogonality via time-domain partitioning.  

For spatial separation, we estimate the angle of arrival (AoA). Because first-order multipath reflections dominate mmWave frequencies, the AoA distribution exhibits sharp peaks. 
We therefore apply a well known forward-backward spatially smoothed MUSIC algorithm based on \cite{Pillai.1989}. The resulting AoAs for both desired and jamming signals enable beamforming-based spatial filtering at the receiver, implemented with a linearly constrained minimum-variance (LCMV) beamformer \cite{Bourgeois.2009}.
When the receiver cannot separate the signals spatially, temporal orthogonality is needed.

\subsection{Jamming classification }
 To differentiate among the attack types listed in Table \ref{tab:jammer_models}, the jammer's fading channel must first be estimated so that channel effects do not bias the classification. A maximum-likelihood estimator is employed for this purpose. After the initial estimate, the channel coefficient is refreshed and stored whenever a jamming waveform is observed. 

For the classification of the DRFM jammer, we propose a stepwise classification process, which employs a novel similarity ratio normalizing cross-correlation by self-correlation (SC) to provide robust detection less sensitive to inherent signal structures. This novel similarity ratio evaluates how the jamming signal resembles  the legitimate signal. First, the self-correlation $ R_{yy}(\tau)$ of the orthogonal legitimate signal is  calculated based on  Eq. (\ref{eq:CC}).
Afterwards the SC maximum value is normalized along the signal duration, and used as reference, i.e.,
\begin{equation}
    SC_{max} = \frac{\max \left( |R_{yy}(\tau)| \right)}{F_{max}}
\end{equation}

Next, the cross-correlation between the orthogonally received jamming signal $\hat{j}$  and the legitimate signal $\hat{y}$ is calculated, given by Eq. (\ref{eq:CC}). 
The maximum cross-correlation $CC_{max}$ value is also normalized. 

Finally, the  similarity ratio between the cross-correlation $CC$ and the self-correlation $SC$ is calculated as:

\begin{equation}
    \mathrm{Sim} = \frac{\max( |R_{\hat{j}\hat{y}}(\tau)|}{\max( |R_{\hat{y}\hat{y}}(\tau)|)} 
\end{equation}

A similarity ratio that surpasses a predefined threshold leads to the classification of the interference as DRFM jamming.  

For amplitude-shifting (AS) and phase-shifting (PS) jammers,  the isolated waveform is first demodulated and correlated with pilot symbols.  If the bit-inversion count surpasses a preset threshold, the jammer is classified as AS or PS according to the active modulation. A similarity ratio is likewise computed; since AS and PS randomly invert parts of the signal, their similarity to the legitimate waveform is significantly lower than that of a DRFM jammer. 


\subsection{Signal, modulation and code adaptation}
Antifragile gains can only be obtained when the legitimate waveform is modified such that the jammer affects none of its information dimensions.  
Signal adaptations therefore focus on preserving the legitimate signal's phase, amplitude, and other dimensions, while treating the jamming waveform as an independent, utilizable resource.
For an AS jammer that perturbs the signal envelope, the system remaps its waveform to an $M$-PSK constellation, thereby removing amplitude dependence. 
With a PS jammer, the transmitter 
adopts ASK whose constellation points reside exclusively in the positive real half-plane, allowing any $180^{\circ}$ phase inversions to be easily corrected. For a DRFM jammer no modulation change is necessary, since the strategy leverages the PSK signal's coherent addition with the jammer's delayed replica.

\label{subsec:Code_adapt} 
 Once the signal is adapted, a new code rate is determined based on the increased SNR value. Therefore, we denote the 
jamming SNR  as  $\text{SNR}_J$ and legitimate SNR as $\text{SNR}_L$:
\begin{equation}
  \mathrm{SNR}_J = \frac{\gamma_{\mathrm{E}}\,\gamma_{\mathrm{J}}}
                          {\gamma_{\mathrm{E}} + \gamma_{\mathrm{J}} + 1},
  \qquad
  \mathrm{SNR}_L = \gamma_L. 
  \label{eq:SNR_jam}
\end{equation}

where $h_L$ denotes the cascaded legitimate channel form the destination to the RIS and $h_E$ and $h_J$ denote the eavesdropping and jamming channel based on Fig.\ref{fig:arch} 
$\begin{aligned}
  \gamma_{\mathrm{E}} &= \tfrac{P_t |h_E|^2}{\sigma_E^2}, \quad
  \gamma_{\mathrm{J}} &= \tfrac{P_j |h_J|^2}{\sigma_J^2}, \quad
  \gamma_{\mathrm{L}} &= \tfrac{P_t |h_L|^2}{\sigma_L^2}.
\end{aligned}$
Since $(n,k)$  Reed-Solomon code can correct up to $t=(n-k)/2$ symbol errors, choosing the optimal code depends on the post-demodulation error profile. Thus, the measured SNR is used to compute the bit error rate (BER) for each adaptive modulation scheme, which is then converted into the residual symbol error rate $P^{\mathrm{RS}}_{\mathrm{res}}$ following the formulation in \cite{Phung.2022}.
\begin{equation}
    P^{\mathrm{RS}}_{\mathrm{rs}} = \frac{n \cdot \mathrm{SER}-t}{n}
\label{eq:RS_selection}
\end{equation}
with $\mathrm{SER} = 1-(1-\mathrm{BER})^{\log_2(M)}$. Using this factor, thresholds for each jamming category can be defined such that $P^{\mathrm{RS}}_{\mathrm{rs}} \leq \Delta < 0$, with $\Delta$ being a negative real number. 

\subsection{Antifragile gain}
\label{subsec:AF_gain}
We denote the throughput in the baseline scenario by $T_{L}$ and under jamming attack by $T_J$ respectively. The optimal Reed-Solomon code is selected according to Eq. \eqref{eq:RS_selection}, which ensures that all symbol errors within each block are corrected. The throughput is then defined based on the code rate $R_C$, bandwidth $B$, modulation order $M$ and SNR index $i$ (Eq. ~\ref{eq:SNR_jam}):
\begin{equation}
  T_i
  = B R_c^{i}\,\log_2(M^{i}), i=L \text{ or } J
  \label{eq:throughput}
\end{equation}

The system demonstrates antifragile gain only if its throughput of the paths with jammers surpasses its baseline, jammer-free throughput. To evaluate this gain across different SNRs, the jamming-to-signal ratio (JSR) is defined. It represents the ratio of jamming power $P_J$ to legitimate signal power $P_L$ between source and destination:
\begin{equation}
    \mathrm{JSR} = 10*\log_{10}(P_J)-10*\log_{10}(P_L)
\label{eq:JSR}
\end{equation}

\section{Numerical Evaluation}
\label{sec:results}

For simulations, we assume that RIS is positioned with respect to the receiver and transmitter at $d_{\mathrm{SR}} = 18~\mathrm{m}$ and $d_{\mathrm{RD}} = 7~\mathrm{m}$, respectively. We set the path-loss exponent to $\delta = 2.7$ and the rate  $\lambda_R = 0.05$. 
The transmitter sends the legitimate signal with a fixed power of $20~\mathrm{dBm}$ on a $28~\text{GHz}$ carrier. Because the receiver's signal-to-noise ratio (SNR) can fluctuate significantly depending on the size of the RIS, the jamming-to-signal ratio (JSR) is defined based on Eq.~\ref{eq:JSR}. This choice confines the jammer output to the interval $0\text{-}40~\text{dBm}$; an SNR-based definition would require substantially higher jammer powers to span the same JSR range at larger baseline SNRs.  
Adaptive modulation and coding (AMC) employs Reed-Solomon codes whose code rates vary from 0.70 (RS$(178,255)$) to 0.94 (RS$(240,255)$) based on Eq. \ref{eq:RS_selection}. The modulation format is selected according to the attack detected, with orders of up to $M=64$.  
Results are reported separately for source-aware and RIS-aware eavesdropping paths, and antifragile gains are emphasized with light blue shading. In all simulations, the channel coefficients were iterated $200$ times, to obtain average values.

\begin{figure}[h!]
\centering
\begin{subfigure}[t]{0.4\textwidth}
\centering
   \includegraphics[scale=0.40]{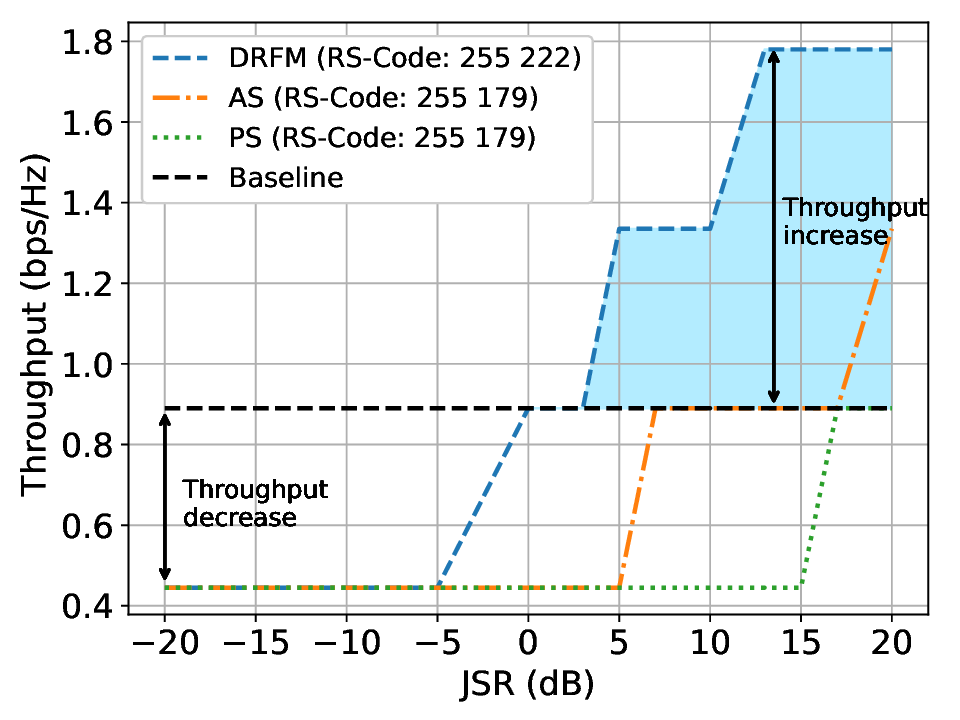}
   \caption{Without orthogonality}
\end{subfigure}
\begin{subfigure}[t]{0.4\textwidth}
\centering
   \includegraphics[scale=0.40]{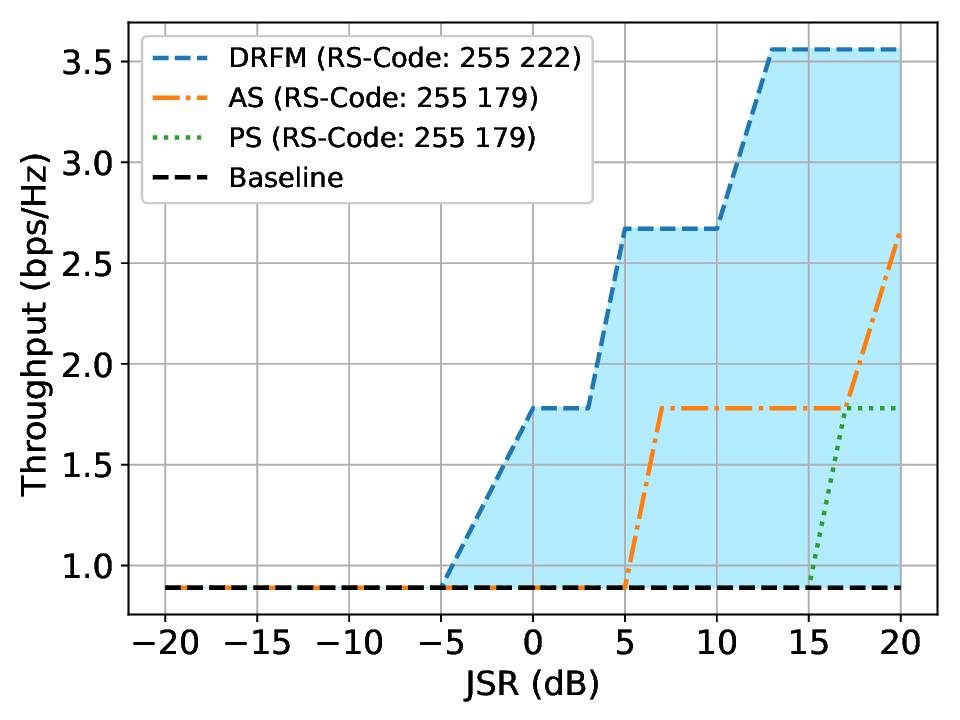}
   \caption{With orthogonality}
\end{subfigure}
\caption{Throughput vs JSR for a fixed coding rate. Baseline SNR = 7~dB}
\label{fig:first}
\end{figure}

\begin{figure}
\centering
\includegraphics[scale=0.32]{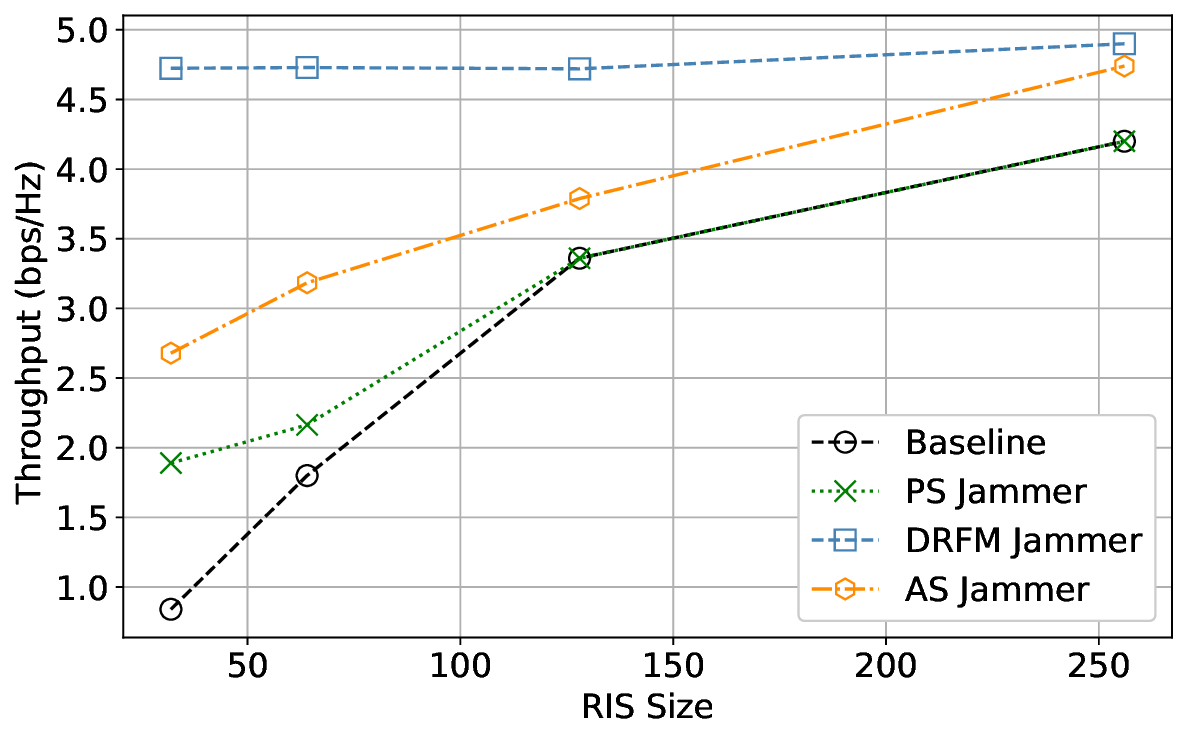}
\caption{Throughput vs JSR for optimal coding rate and varying RIS sizes.}
\label{fig:third}
\end{figure}

Figure \ref{fig:first} compares the three attacks with a fixed code rate.  
 Without spatial orthogonality the jammer and desired signal overlap for half the frame; the transmitter shortens its burst, 
 reducing payload and throughput (Eq.~\ref{eq:throughput}). Thus,  antifragile gains appear only after raising the modulation order, i.e., at JSR = 3 dB for DRFM and 15 dB for AS.  
 With full orthogonality 
 and a higher baseline SNR, throughput exceeds the jammer, free reference from about -5 dB, DRFM allowing for the largest boost due to its high mutual information with the desired waveform.
Figure \ref{fig:third} scales the RIS size. As the surface grows, the legitimate SNR increases faster than the fixed-power jammer can follow, so antifragile benefits fade; the PS attack offers no gain beyond 128 elements. Across the source-aware scenario DRFM delivers the highest data rates, followed by AS and PS, consistent with mutual-information limits under fixed jammer power and position.

\begin{figure}[h!]
\centering
\includegraphics[scale=0.4]{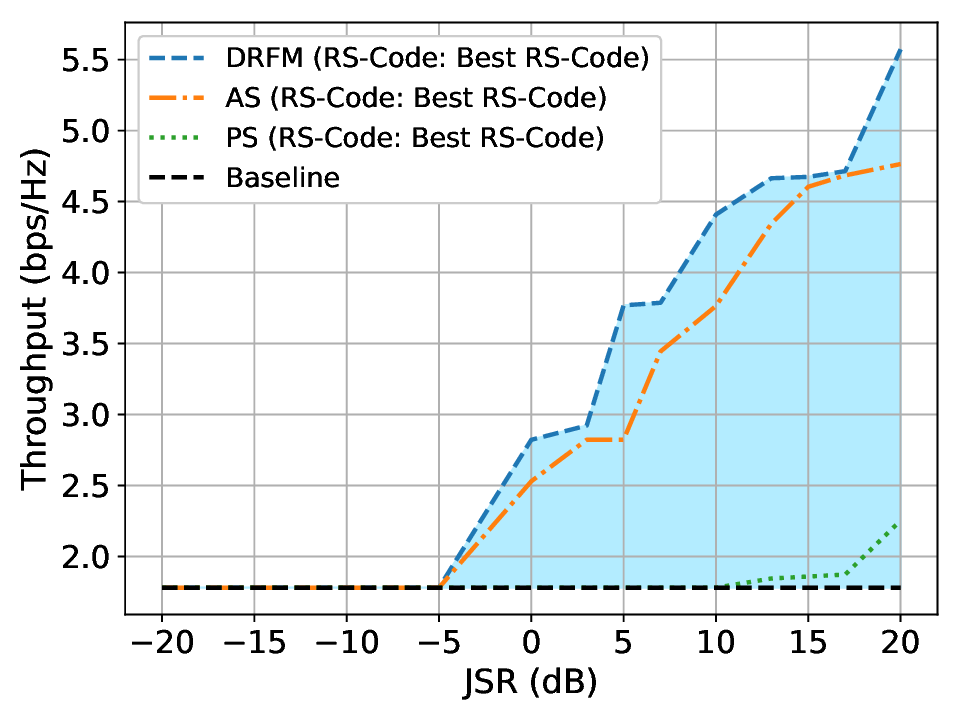}
\caption{Throughput vs JSR for optimal selected coding rate. Baseline SNR = 10~dB}
\label{fig:fith}
\end{figure}

For RIS-aware eavesdropping path, we keep the RIS location, code rate, and modulation order identical to source-aware path.  
Figure \ref{fig:fith} presents the reactive-jammer results under the same baseline SNR and adaptive coding settings used in Fig. \ref{fig:third}. Here, every jammer achieves a noticeably higher throughput. 
This outcome is as expected because the adversary receives the RIS-reflected waveform and thereby benefits from its beam-forming gain.  Consequently, the effective jamming SNR scales more rapidly as JSR grows.

 \section{Conclusion}\label{sec:conclusion}

This paper presented for the first time a cross-layer network integration approach of RIS-assisted link antifragility under jamming attacks.  
The results confirm that RIS, when coupled with an antifragile design philosophy, can convert hostile interference into a throughput resource. The proposed antifragile behavior is the basis for cross-layer integration with network control mechanisms such as adaptive routing and link scheduling.
This insight opens up a new design dimension for resilient wireless networks, where deliberate or accidental interference may be exploited rather than merely avoided.




\section*{Acknowledgment}
This work was partially supported by the project MARE (Grant Agreement No 101191436), under  Smart Networks and Services Joint Undertaking (SNS JU) EU Horizon program.
\bibliographystyle{IEEEtran}
\bibliography{mybib}

\end{document}